\documentclass[twocolumn,shortnote]{jpsj3}
\usepackage{txfonts}

\title{
Absence of anomalous negative lattice-expansion 
for polycrystalline sample of Tb$_2$Ti$_2$O$_7$
}

\author{
\name{Kazuki \surname{Goto}}, 
\name{Hiroshi \surname{Takatsu}}, 
\name{Tomohiro \surname{Taniguchi}}, and 
\name{Hiroaki \surname{Kadowaki}} }

\inst{
Department of Physics, Tokyo Metropolitan University, 
Hachioji-shi, Tokyo 192-0397 }

\recdate{\today}

\kword{X-ray diffraction, spin liquid, lattice expansion}

\begin{document}
\maketitle

Magnetic systems with geometric frustration, 
a prototype of which is antiferromagnetically coupled 
Ising spins on a triangle, have been intensively studied 
experimentally and theoretically for decades \cite{Lacroix11}. 
Spin systems on networks of triangles or tetrahedra, 
such as triangular \cite{Wannier50}, kagom\'{e} \cite{Shyozi51}, 
and pyrochlore \cite{Gardner10} lattices, play major roles 
in these studies. 
Subjects that have fascinated many investigators in recent 
years are classical and quantum spin-liquid states \cite{Lee08,Balents10}, 
where conventional long-range order (LRO) is suppressed 
to very low temperatures. 
Quantum spin-liquids in particular have been 
challenging both theoretically and experimentally 
since the proposal of the resonating valence-bond state \cite{Anderson73}. 
The spin ice materials, R$_2$T$_2$O$_7$ (R = Dy, Ho; T = Ti, Sn), are
the well-known classical examples \cite{Bramwell01}, 
while other experimental candidates found recently are awaiting further studies \cite{Balents10}.

Among frustrated magnetic pyrochlore oxides \cite{Gardner10}, 
Tb$_2$Ti$_2$O$_7$ has attracted much attention 
because it does not show any conventional LRO down to 50 mK 
and remains in a dynamic spin-liquid state \cite{Gardner99s,Gardner03s}. 
Theoretical considerations of the crystal-field states 
of Tb$^{3+}$ and exchange and dipolar interactions of 
the system \cite{Gingras00s,Enjalran04,Kao03} 
showed that it should undergo a transition into 
a conventional magnetic LRO state at about $T \sim 1.8$ K within a random 
phase approximation \cite{Kao03}. 
The puzzling origin of the spin-liquid state has been 
in debate \cite{Gardner10}. 
An interesting scenario to explain the spin-liquid state is the theoretical 
proposal of a quantum spin-ice state \cite{Molavian07}. 

Recently we have pointed out \cite{Takatsu11} that 
several experimental puzzles of Tb$_2$Ti$_2$O$_7$ 
originate from difficulty of controlling quality of 
crystalline samples \cite{Chapuis09URL,Hamaguchi04,Gingras00s}, 
and that experiments on polycrystalline samples provided 
consistent evidence of the spin-liquid state. 
In addition, we have performed 
neutron scattering and specific heat experiments 
on polycrystalline samples \cite{Takatsu11}, 
and have found that there should be a overlooked 
cross-over temperature at $T \sim 0.4$ K 
for the polycrystalline sample used in the seminal 
work \cite{Gardner99s,Gardner03s} of the spin-liquid state of Tb$_2$Ti$_2$O$_7$. 
The spin-liquid ground-state does have quantum spin-fluctuations \cite{Takatsu11}. 

The purpose of the present study is to check whether 
a well-characterized polycrystalline sample shows 
the anomalous negative thermal-expansion below 20~K, 
which was intriguingly reported for a crystal sample \cite{Ruff07}. 
This anomalous behavior was observed simultaneously with 
the broadening of the Bragg peaks, which could be attributed 
to cubic-tetragonal structural fluctuations brought about by 
a magneto-elastic coupling \cite{Ruff07,Mamsurova86,Nakanishi11}. 
In connection with these anomalies, another mechanism of 
the spin-liquid state was proposed, where the crystal field 
ground doublet becomes two singlets owing to the tetragonal 
distortion \cite{Bonville11s}. 
We have measured temperature dependence 
of the lattice constant $a(T)$ on a polycrystalline sample very precisely 
down to $T=4$~K, but did not observe the negative 
thermal-expansion. 

A polycrystalline sample of Tb$_2$Ti$_2$O$_7$ was 
prepared by the standard solid-state reaction \cite{Gardner99s}. 
Stoichiometric mixtures of Tb$_2$O$_3$ and TiO$_2$ 
were heated in air at 1350~$^\circ$C for several days with 
intermittent grindings to ensure a complete reaction. 
It was ground into powder and annealed in air 
at 800$^\circ$C for one day. 
The specific heat of this sample is the same as 
that reported in ref.~\citen{Takatsu11}. 
X-ray powder-diffraction experiments were carried out 
using a RIGAKU-SmartLab powder diffractometer 
equipped with a Cu K$_{\alpha 1}$ monochromator. 
The sample was mounted in a closed-cycle He-gas refrigerator. 

\begin{figure*}[t]
\begin{center}
\includegraphics[width=15.0cm,clip]{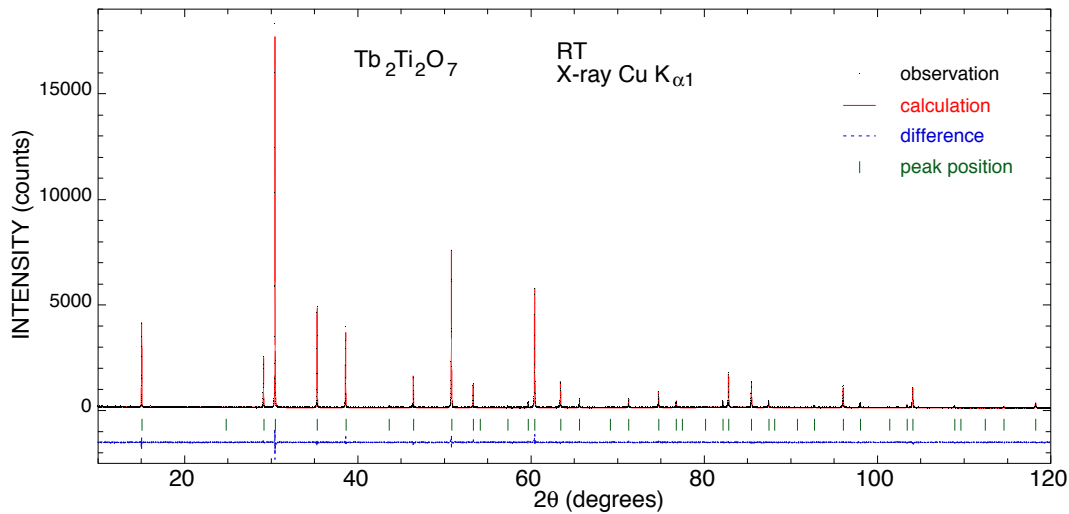}
\end{center}
\caption{
(Color online) 
X-ray powder-diffraction pattern of Tb$_2$Ti$_2$O$_7$ at 
room temperature. 
Observed and calculated patterns are denoted by 
closed circles and a solid line, respectively. 
Their difference is plotted in lower part by a dashed line.
Vertical bars stand for positions of Bragg reflections.
}
\label{X-ray1004}
\end{figure*}

To study the sample quality, we analyzed an X-ray 
powder-diffraction pattern taken at a room temperature (RT) 
by using the Rietveld profile refinement 
program RIETAN-97 \cite{izumi}. 
It was shown that the powder pattern is consistent with the 
fully ordered cubic pyrochlore structure. 
In Fig.~\ref{X-ray1004} we show the diffraction 
pattern and its analysis. 
The pyrochlore structure belongs to the space group $Fd\bar{3}m$, 
and constituent atoms occupy the sites of 
16$d$(Tb), 16$c$(Ti), 48$f$(O), and 8$b$(O'). 
The refined structure parameters are listed in 
Table~\ref{tab.str1004}, 
which agree well with those obtained by the previous neutron 
powder-diffraction experiments \cite{Han04}. 
The final $R$ factors of the refinement are 
$R_{\text{wp}} = 8.1$ \% and $R_{\text{e}} = 7.5$ \% ($S = 1.1$), 
implying that the quality of the fit is very good. 
\begin{table}
\caption{
Refined structure parameter at RT 
using X-ray powder-diffraction data of Fig.~\ref{X-ray1004}. 
Number in parenthesis is standard deviation of the last digit.
Values of $B$ are fixed for 48$f$(O), and 8$b$(O').
}
\label{tab.str1004}
\begin{center}
%\begin{tabular}{dddddd}
\begin{tabular}{cccccc}
\hline
\multicolumn{3}{c}{$Fd\bar{3}m$ (No. 227)} 
& \multicolumn{3}{c}{$a=$ 10.152(2) \AA} \\ 
Atom & Site  & $x$ & $y$ & $z$ & $B$~${\rm \AA}^2$ \\
\hline
Tb & 16$d$ & 1/2 & 1/2  & 1/2 & 0.5(1) \\
Ti & 16$c$ & 0   & 0    & 0   & 0.6(2) \\
O  & 48$f$ & 0.324(3) & 1/8  & 1/8 & 0.4 \\
O' &  8$b$ & 3/8       & 3/8  & 3/8 & 0.4 \\
\hline
\end{tabular}
\end{center}
\end{table}
\begin{figure}
\begin{center}
\includegraphics[width=8.5cm,clip]{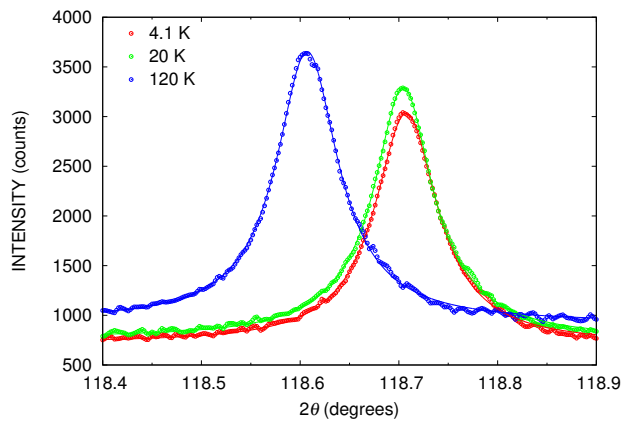}
\end{center}
\caption{
(Color online) 
Temperature dependence of $\theta$-$2\theta$ scans 
around (8,8,0) reflection using the powder sample.
}
\label{scan880}
\end{figure}
In order to measure $T$ dependence 
of the lattice constant $a(T)$, we performed 
$\theta$-$2\theta$ scans around the (8,8,0) reflection 
in a temperature range between 4 and 120~K. 
In Fig.~\ref{scan880} we show the scans plotted 
as a function of $2\theta$. 
One can see that the peak positions are 
determined within a very small experimental error of 
0.001 degrees. 
By converting the peak position to the lattice constant, 
we show $\Delta a(T)/a(20\text{K})$, where 
$\Delta a(T) = a(T)-a(20\text{K})$, in Fig.~\ref{LC} 
together with the previous experimental data of 
a crystal sample (Fig. 4 of ref.~\citen{Ruff07}). 
The powder sample shows the same $T$ dependence 
of $\Delta a(T)/a(20\text{K})$ as the crystal above 20~K. 
However below 20 K the powder sample 
shows only normal positive thermal-expansion. 
We speculate that unknown control parameters of quality of 
crystals affects the anomalous negative thermal-expansion of 
the crystalline sample \cite{Ruff07}. 
The present result implies that 
the spin-liquid state established by experiments 
on polycrystalline samples \cite{Gardner99s,Gardner03s,Takatsu11} 
cannot be simply explained by the two-singlets mechanism 
based on the tetragonal distortion \cite{Bonville11s}.
\begin{figure}
\begin{center}
\includegraphics[width=8.5cm,clip]{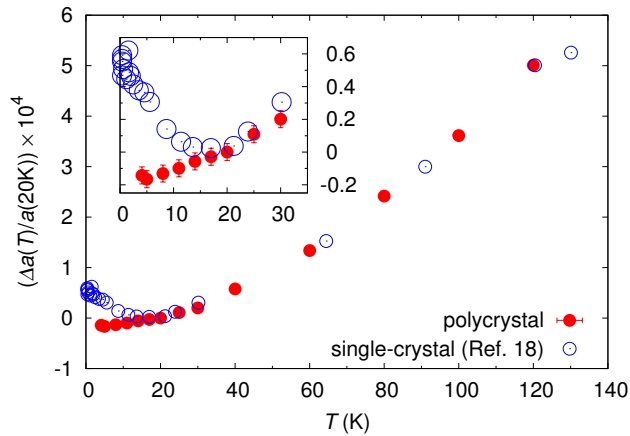}
\end{center}
\caption{
(Color online) 
Temperature dependence of the lattice constant normalized at 
$T=20$~K of the powder sample and the previous measurement 
using a crystal sample \cite{Ruff07}. The inset shows 
normal positive and anomalous negative 
lattice-expansion below $T<20$~K of the powder and 
crystal samples, respectively.
}
\label{LC}
\end{figure}

\begin{acknowledgments}
We thank R. Higashinaka, and K. Matsuhira 
for useful discussions. 
This work was supported by 
KAKENHI on Priority Areas ``Novel States of Matter Induced by Frustration''. 
\end{acknowledgments}

\bibliographystyle{jpsj}
\bibliography{TTO_G}

\end{document}